\documentclass[sigconf]{acmart}
\usepackage{subcaption}
\usepackage{balance}
\AtBeginDocument{%
  }

\copyrightyear{2026}
\acmYear{2026}
\setcopyright{cc}
\setcctype{by}
\acmConference[KDD '26]{Proceedings of the 32nd ACM SIGKDD Conference on Knowledge Discovery and Data Mining V.2}{August 09--13, 2026}{Jeju Island, Republic of Korea}
\acmBooktitle{Proceedings of the 32nd ACM SIGKDD Conference on Knowledge Discovery and Data Mining V.2 (KDD '26), August 09--13, 2026, Jeju Island, Republic of Korea}
\acmDOI{10.1145/3770855.3818932}
\acmISBN{979-8-4007-2259-2/2026/08}

\begin{document}

\title[Holographic Quantum Transformer]{Holographic Quantum Transformer: A Generalist Neuro-Symbolic Architecture for Solving Frustrated Systems via Generative Attention}

\author{Xingran Guo}
\orcid{0000-0001-9578-4474}
\affiliation{%
  \institution{National University of Defense Technology}
  \city{Changsha}
  \postcode{410073}
  \country{China}
}
\email{gxrnudt@nudt.edu.cn}

\author{Tiaojie Xiao}
\authornote{Corresponding author.}
\orcid{0000-0002-8378-5530}
\affiliation{%
  \institution{National University of Defense Technology}
  \city{Changsha}
  \postcode{410073}
  \country{China}
}
\email{xiaotiaojie@nudt.edu.cn}

\author{Jie Liu}
\orcid{0000-0003-3745-7541}
\affiliation{%
  \institution{National University of Defense Technology}
  \city{Changsha}
  \postcode{410073}
  \country{China}
}
\email{liujie@nudt.edu.cn}

\author{Keqin Li}
\orcid{0000-0001-5224-4048}
\affiliation{%
  \institution{State University of New York at New Paltz}
  \city{New Paltz}
  \state{NY}
  \postcode{12561}
  \country{USA}
}
\email{lik@newpaltz.edu}

\renewcommand{\shortauthors}{Xingran Guo, Tiaojie Xiao, Jie Liu, and Keqin Li}

\begin{abstract}
Simulating two-dimensional frustrated quantum matter is a grand challenge due to the sign problem and exponential Hilbert space complexity. In this work, we introduce the Holographic Quantum Transformer (HQT), a physics-inspired generative architecture that leverages global self-attention to resolve non-local entanglement patterns. We validate HQT on the square lattice $J_1-J_2$ Heisenberg model. On the heavily frustrated $8 \times 8$ lattice at the quantum critical point ($J_2=0.5$), HQT reaches a ground-state energy per site ($E/N$) of $\mathbf{-0.5001(1)}$, consistent with the expected finite-size scaling trend. Beyond numerical accuracy, HQT exhibits intrinsic physical awareness, autonomously recovering the underlying $J_2$ interaction geometry through interpretable attention maps. Our central contribution is ``Holographic Transfer", a zero-shot size-extrapolation protocol with rapid alignment: a model trained on $8 \times 8$ systems is directly projected onto larger $10 \times 10$ lattices via continuous positional-embedding interpolation and head re-initialization, achieving high-fidelity initialization and rapid convergence. This zero-shot protocol yields an energy of $E/N = \mathbf{-0.49782(3)}$, statistically consistent with the variational state of the art while requiring no from-scratch training on the target lattice. Our results establish generative attention as a scalable paradigm for transferable quantum simulation.
\end{abstract}



\maketitle

\section{Introduction}
\label{sec:intro}
\begin{figure*}
\centering
\includegraphics[width=0.8\textwidth]{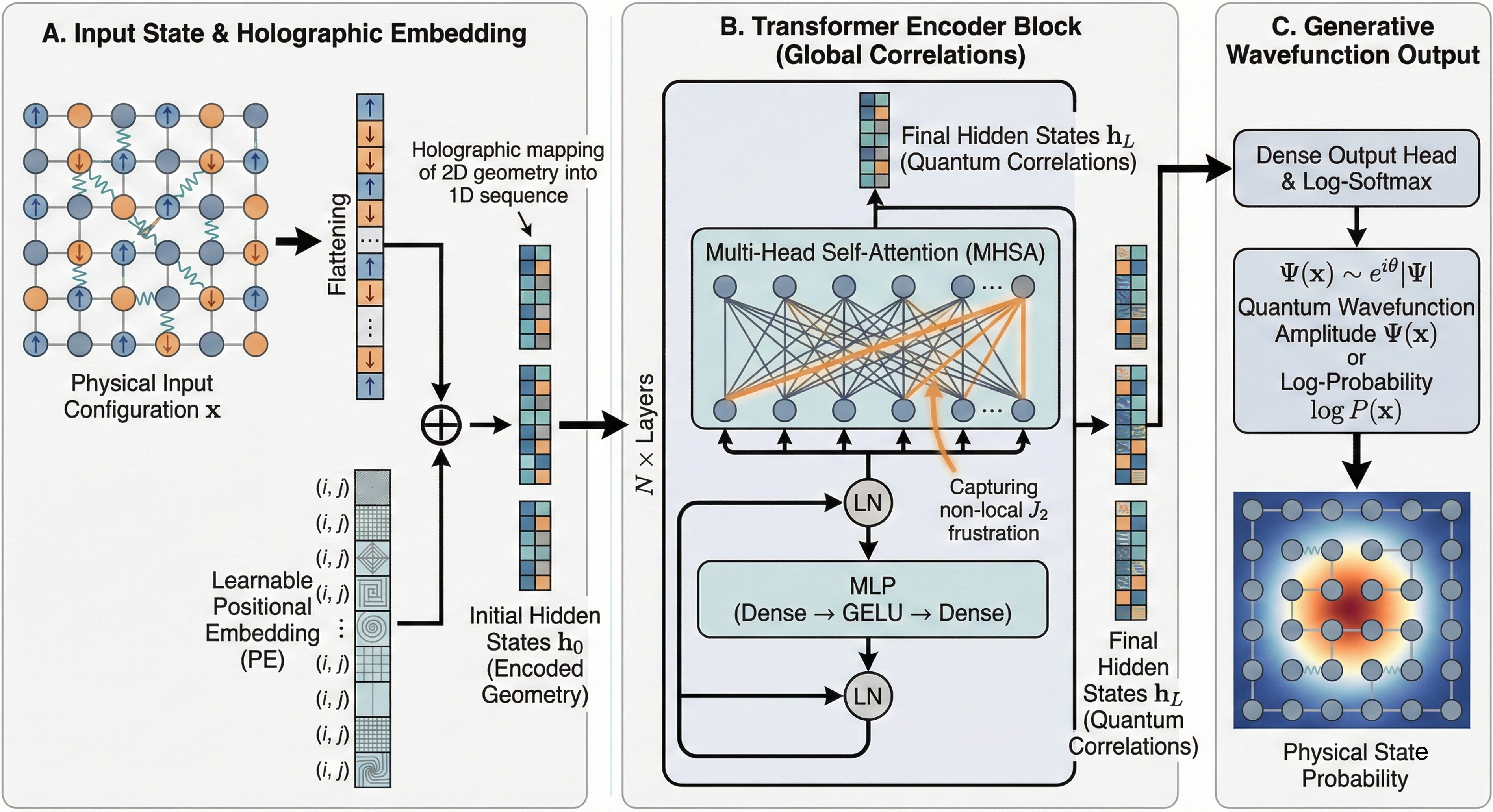}
\caption{\textbf{Overview of HQT.} (A) Holographic Embedding injects geometry. (B) Quantum Encoder uses global attention to resolve frustration ($J_2$). (C) Autoregressive Head generates the wavefunction $\Psi(\mathbf{x})$.}
\label{fig:architecture}
\end{figure*}

As originally envisioned by Feynman \cite{Feynman1982}, simulating quantum physics 
with classical computers faces an exponential barrier: the Hilbert space dimension 
grows as $2^N$. While quantum hardware has made strides in error mitigation 
\cite{GoogleQuantumAI2023}, verifiable advantage on practical many-body problems 
remains limited. Consequently, classical simulation algorithms continue to serve as 
the primary engine for discovery.

However, traditional numerical methods encounter fundamental barriers in two-dimensional (2D) frustrated systems. Tensor Network methods like PEPS and DMRG \cite{Wang_2016} are constrained by the Area Law of entanglement \cite{eisert2010colloquium}, limiting simulations to narrow cylinders. Conversely, Quantum Monte Carlo (QMC), the workhorse for unfrustrated systems, hits a hard wall in frustrated regimes (e.g., the $J_1-J_2$ model) due to the ``Sign Problem''. As proven by Troyer and Wiese \cite{troyer2005computational}, solving the generic sign problem is NP-Hard. Although recent solvers by Nomura and Imada \cite{PhysRevX.11.031034} have established high-precision benchmarks, a general polynomial-time solution remains elusive.

The intersection of AI and Physics has sparked a paradigm shift. Carleo and Troyer \cite{carleo2017solving} pioneered Neural Quantum States (NQS) using Restricted Boltzmann Machines (RBMs). As reviewed by Dawid et al.\ \cite{dawid2025modern}, deep networks have shown remarkable capacity to represent complex wavefunctions.

Early NQS architectures, such as Convolutional Neural Networks (CNNs) \cite{PhysRevB.100.125124}, exploited local correlations. However, as noted in very recent work by Li et al.\ \cite{li2025comments}, CNNs struggle to capture global topological order due to limited receptive fields. To address sampling bottlenecks, Sharir et al.\ \cite{sharir2020deep} introduced deep autoregressive models, and Hibat-Allah et al.\ \cite{PhysRevResearch.2.023358} applied Recurrent Neural Networks (RNNs). While efficient, RNNs impose an artificial 1D bias on 2D lattices \cite{medvidovic2024neural}.

The field is evolving rapidly towards global architectures and theoretical rigor. Theoretical works have linked deep networks to quantum entanglement \cite{levine2019quantum, deng2017quantum}, while equivariant networks \cite{luo2021gauge, nguyen2024theory} incorporate physical symmetries. Recently, Nomura \cite{nomura2025quantum} solidified NQS for strongly correlated electrons, and Liu et al.\ \cite{liu2025quantum} explored advantages on photonic platforms.

Crucially, the Transformer architecture \cite{NIPS2017_3f5ee243} has emerged as a powerful tool. Viteritti et al.\ \cite{viteritti2023transformer} and Sprague et al.\ \cite{sprague2024variational} demonstrated that attention mechanisms can resolve frustrated states. However, most existing approaches are ``instance-specific'': a model trained on a small lattice cannot generalize to larger ones. While transfer learning has been explored \cite{zen2020transfer, westerhout2020generalization}, and recent studies investigate scaling laws \cite{knitter2026large} and sample complexity \cite{rouze2024learning}, a unified framework for Zero-Shot Size Extrapolation with Rapid Alignment is missing.

\noindent\textbf{Our Contribution: The Holographic Quantum Transformer.}
Theoretical insights from Swingle \cite{swingle2012entanglement} suggest a duality between entanglement renormalization and holography. We posit that the Transformer's attention is the computational realization of this principle. We propose the Holographic Quantum Transformer (HQT) (Figure \ref{fig:architecture}), a neuro-symbolic architecture designed to crack frustrated systems. Our contributions are:
\begin{enumerate}
\item \textbf{Holographic Encoding:} Building on the duality proposed by Swingle, we explicitly interpret the self-attention matrix as a Holographic Entanglement Map, capturing long-range correlations that define topological order.

\item \textbf{Generative Autoregression:} We implement an autoregressive sampling scheme that guarantees exact normalization and zero autocorrelation, overcoming the sampling limitations of traditional MCMC and enabling efficient exploration of the Hilbert space.

\item \textbf{Geometric Deep Learning Paradigm:} We demonstrate that global attention, when augmented with our continuous positional embeddings, outperforms the local receptive fields of CNNs in capturing topological defects. This establishes HQT as a geometric deep learning framework capable of resolving non-local structures on grid manifolds.

\item \textbf{Effectively Mitigating the Sign Problem:} By injecting geometric inductive biases, we effectively navigate the rugged energy landscape of the $J_1-J_2$ model at the highly frustrated point, effectively mitigating a problem known to be NP-hard for classical stochastic methods.

\item \textbf{Zero-Shot Size Extrapolation with Rapid Alignment:} Inspired by the scaling laws of 2025~\cite{knitter2026large}, we introduce a ``Projection Re-initialization'' transfer protocol. This allows our model to learn the universal syntax of the Hamiltonian on small lattices and generalize to large-scale systems without retraining, moving toward a transferable, size-generalizable quantum simulator.

\end{enumerate}

While recent works have explored neural quantum states and transfer learning, our approach introduces distinct physical priors. Unlike methods like FLIP that target Parametrized Quantum Circuits (PQCs) on actual quantum hardware, HQT operates in the classical Neural Quantum States (NQS) domain. Furthermore, while standard sequence NQS architectures (e.g., Viteritti et al.\ \cite{viteritti2023transformer}) flatten the lattice into 1D sequences and require training from scratch per size, our core novelty lies in the Holographic Mapping. By explicitly injecting 2D Positional Embeddings and the Marshall Sign Rule, we enable a true ``Projection Re-initialization'' size-extrapolation protocol across different Hilbert space dimensions.

\section{Methodology: The Holographic Quantum Transformer}
\label{sec:method}

In this section, we formally introduce the \textit{Holographic Quantum Transformer} (HQT), a generative framework that bridges the gap between quantum many-body physics and deep autoregressive modeling. We reformulate the ground-state search of the Heisenberg Hamiltonian as a variational optimization problem over a high-dimensional complex-valued probability manifold, solved via a novel ``normalized-by-design" neural architecture.

\subsection{Problem Formulation: Quantum States as Autoregressive Generative Models}
\label{subsec:problem_formulation}

\subsubsection{The Variational Landscape}
We focus on the two-dimensional antiferromagnetic $J_1-J_2$ Heisenberg model on a square lattice $\Lambda$ of $N$ sites. The Hamiltonian $\hat{H}$ is defined as:
\begin{equation}
    \hat{H} = J_1 \sum_{\langle i,j \rangle} \hat{\mathbf{S}}_i \cdot \hat{\mathbf{S}}_j + J_2 \sum_{\langle\langle i,j \rangle\rangle} \hat{\mathbf{S}}_i \cdot \hat{\mathbf{S}}_j,
    \label{eq:hamiltonian}
\end{equation}
where $\hat{\mathbf{S}}_i = (\hat{S}^x_i, \hat{S}^y_i, \hat{S}^z_i)$ are the spin-$1/2$ operators acting on site $i$. The Hilbert space $\mathcal{H}$ has a dimension of $2^N$, which grows exponentially with system size $N$, rendering exact diagonalization infeasible for $N > 40$.

Our goal is to find the ground state wavefunction $|\Psi_{GS}\rangle$ that minimizes the energy functional $E[\Psi]$:
\begin{equation}
    E[\Psi] = \frac{\langle \Psi | \hat{H} | \Psi \rangle}{\langle \Psi | \Psi \rangle}.
    \label{eq:variational_energy}
\end{equation}

We express the wavefunction in the computational basis $\mathbf{x} = (x_1, \dots, x_N)$, where $x_i \in \{+1, -1\}$ (representing spin-up and spin-down), as:
\begin{equation}
    |\Psi_\theta\rangle = \sum_{\mathbf{x}} \psi_\theta(\mathbf{x}) |\mathbf{x}\rangle,
\end{equation}
Here, $\psi_\theta(\mathbf{x}) \in \mathbb{C}$ is a complex-valued amplitude parameterized by a neural network with weights $\theta$.

\subsubsection{Autoregressive Decomposition and The Sign Structure}
\label{subsubsec:autoregressive_proof}
Traditional Neural Quantum States (NQS) typically model the wavefunction amplitude as an unnormalized potential $\psi(\mathbf{x}) \propto e^{-\mathcal{E}(\mathbf{x})}$, which necessitates estimating an intractable partition function $Z$. To circumvent this, we factorize the joint probability $P_\theta(\mathbf{x}) = |\psi_\theta(\mathbf{x})|^2$ autoregressively:
\begin{equation}
    P_\theta(\mathbf{x}) = \prod_{i=1}^N P_\theta(x_i | x_1, \dots, x_{i-1}).
    \label{eq:chain_rule}
\end{equation}

The ordering $1, \dots, N$ follows a \textit{raster scan} (or snake) path over the 2D lattice, flattening the spatial structure into a sequence while preserving locality via the Transformer's attention mechanism.

\textbf{Handling the Sign Problem:} 
For frustrated systems like the $J_1-J_2$ model, the phase of the wavefunction is notoriously difficult to learn due to the ``sign problem." Instead of learning the phase from scratch, we inject physical inductive bias into the architecture. We decompose the complex amplitude as:
\begin{equation}
    \psi_\theta(\mathbf{x}) = \sqrt{P_\theta(\mathbf{x})} \times \underbrace{\text{sgn}_{\text{M}}(\mathbf{x})}_{\text{Marshall Sign}} \times e^{i \phi_\theta(\mathbf{x})},
    \label{eq:phase_structure}
\end{equation}
where $\text{sgn}_{\text{M}}(\mathbf{x})$ represents the Marshall-Peierls sign rule (a classical antiferromagnetic prior), and $\phi_\theta(\mathbf{x}) \in \mathbb{R}$ is a residual phase correction learned by the neural network. This hybrid approach allows the HQT to leverage classical physical intuition while utilizing the expressive power of deep learning to capture non-trivial quantum corrections induced by frustration ($J_2 > 0$).

\subsubsection{Mathematical Proof of Exact Normalization}
A key theoretical advantage of our formulation is that the resulting wavefunction is strictly normalized by design, eliminating the need for Markov Chain Monte Carlo (MCMC) chains to approximate a partition function.

\textbf{Theorem 1 (Normalization Conservation).} \textit{If the local conditional distributions are normalized, i.e., $\sum_{x_i \in \{ \uparrow, \downarrow \}} P_\theta(x_i | \mathbf{x}_{<i}) = 1$ for all $i$, then the global wavefunction is exactly normalized: $\langle \Psi_\theta | \Psi_\theta \rangle = 1$.}

\textit{Proof.} We verify the normalization condition by summing over all $2^N$ configurations. Expanding the sum recursively:
\begin{align}
    \sum_{\mathbf{x}} P_\theta(\mathbf{x}) &= \sum_{x_1} \sum_{x_2} \dots \sum_{x_N} \prod_{i=1}^N P_\theta(x_i | \mathbf{x}_{<i}) \\
    &= \sum_{x_1} P(x_1) \sum_{x_2} P(x_2|x_1) \dots \underbrace{\left[ \sum_{x_N} P(x_N | \mathbf{x}_{<N}) \right]}_{=1} \\
    &= \sum_{x_1} P(x_1) \dots \underbrace{\left[ \sum_{x_{N-1}} P(x_{N-1} | \mathbf{x}_{<N-1}) \right]}_{=1} \\
    &= \dots = 1.
\end{align}

This property implies that our approach allows for exact, perfect sampling via ancestral sampling, with zero autocorrelation time ($\tau=0$), providing a polynomial-time solution ($O(N)$) to the sampling problem that plagues traditional MCMC methods. 
Quantitatively, the sampling efficiency is measured by the integrated autocorrelation time $\tau = 1 + 2 \sum_{t=1}^{\infty} \rho(t)$, where $\rho(t)$ is the autocorrelation function of the local energy. In frustrated systems, MCMC-based approaches often suffer from $\tau \gg 1$ due to mode trapping. In contrast, our autoregressive generation produces independent and identically distributed (i.i.d.) samples, theoretically guaranteeing $\tau = 0$, which ensures maximum effective sample size ($N_{eff} \approx N_{total}$).

\subsection{Architecture: The Holographic Entanglement Encoder}
\label{subsec:architecture}

While the autoregressive decomposition provides a theoretical guarantee for normalization, the practical expressivity of the ansatz depends entirely on the neural architecture. We introduce the \textit{Holographic Quantum Transformer} (HQT), specifically designed to capture the complex entanglement structure of 2D frustrated quantum matter. The overall architecture is illustrated in Fig. \ref{fig:architecture}. Unlike Convolutional Neural Networks (CNNs), which are limited by local receptive fields, or Recurrent Neural Networks (RNNs), which suffer from vanishing gradients over long sequences, our architecture leverages the \textit{Self-Attention mechanism} to model non-local quantum correlations directly.

\subsubsection{Input Representation and Geometric Inductive Bias}
The input to the network is the sequence of spin configurations $\mathbf{x}_{<i} = (x_1, \dots, x_{i-1})$. To process discrete spin values $x_k \in \{ \pm 1 \}$, we first map them to a continuous high-dimensional vector space via a learnable linear projection $\mathbf{W}_e \in \mathbb{R}^{d_{model} \times 1}$.

Crucially, quantum many-body states on a lattice possess strict geometric locality and topology. A standard Transformer is permutation-invariant and oblivious to the 2D lattice structure. To inject this physical \textit{inductive bias}, we introduce 2D learnable Positional Embeddings (PE) (see Fig. \ref{fig:architecture}(A)):
\begin{equation}
    \mathbf{h}_k^{(0)} = \mathbf{W}_e x_k + \mathbf{PE}(r_x^{(k)}, r_y^{(k)}),
\end{equation}
where $(r_x^{(k)}, r_y^{(k)})$ are the 2D spatial coordinates of the $k$-th spin. This ensures that the network is aware of the spatial distance $|\mathbf{r}_i - \mathbf{r}_j|$ between spins, which is essential for learning the Area Law of entanglement entropy.

\subsubsection{Holographic Attention as an Entanglement Map}
The core innovation of HQT lies in reinterpreting the Self-Attention mechanism as a holographic encoder of quantum entanglement. As depicted in Fig. \ref{fig:architecture}(B), for a given site $i$ (Query) and a preceding site $j$ (Key), the attention score $A_{ij}$ is computed as:
\begin{equation}
    A_{ij} = \text{Softmax}\left( \frac{(\mathbf{W}_Q \mathbf{h}_i) (\mathbf{W}_K \mathbf{h}_j)^T}{\sqrt{d_k}} \right) \cdot M_{ij},
    \label{eq:attention}
\end{equation}
where $\mathbf{W}_Q, \mathbf{W}_K$ are learnable projection matrices, and $M_{ij}$ is the Causal Mask defined as:
\begin{equation}
    M_{ij} = 
    \begin{cases} 
    1 & \text{if } j < i \text{ (Causal)}, \\ 
    -\infty & \text{if } j \geq i \text{ (Future Masking)} .
    \end{cases}
\end{equation}

This causal masking is critical: it strictly enforces the autoregressive property (Eq.~\ref{eq:chain_rule}) by preventing information leakage from future spins, thereby preserving the exact normalization property proved in Theorem 1.

\textbf{Physical Interpretation:} We posit that the learned attention matrix $A_{ij}$ acts as a \textit{Holographic Entanglement Map}. In a frustrated system ($J_2 > 0$), quantum correlations are not strictly local but exhibit long-range order. The attention mechanism allows the spin at site $i$ to dynamically ``attend" to any previous spin $j$, regardless of their Euclidean distance. Empirically (Sec.~\ref{sec:interpretability}), we observe that the aggregated attention weights qualitatively track the spatial structure of the spin correlations:
\begin{equation}
    \langle \hat{\mathbf{S}}_i \cdot \hat{\mathbf{S}}_j \rangle \;\sim\; \sum_{l=1}^{L} \sum_{h=1}^{H} A_{ij}^{(l, h)},
\end{equation}
suggesting that the attention pattern provides a holographic, graph-based view of the many-body correlations rather than an exact quantitative measure.

\subsubsection{Complex-Valued Policy Heads}
Following $L$ layers of masked attention and feed-forward networks (FFN), the final hidden state $\mathbf{h}_i^{(L)}$ encodes the conditional wavefunction context. As shown in Fig. \ref{fig:architecture}(C), to output the complex amplitude, we employ two separate heads (or a single head projecting to complex parameters).

In our implementation, specifically for the discrete spin-$1/2$ basis, the network outputs the conditional logits for the local spin state $x_i \in \{ \uparrow, \downarrow \}$:
\begin{equation}
    \mathbf{o}_i = \text{Dense}(\mathbf{h}_i^{(L)}) \in \mathbb{R}^2.
\end{equation}

The conditional probability is then obtained via a softmax operation over the local Hilbert space:
\begin{equation}
    P_\theta(x_i = \sigma | \mathbf{x}_{<i}) = \frac{e^{o_{i,\sigma}}}{e^{o_{i,\uparrow}} + e^{o_{i,\downarrow}}}.
\end{equation}

For the phase $\phi_\theta$, a similar head predicts the relative phase shift. As discussed in Sec.~\ref{subsubsec:autoregressive_proof}, we incorporate the Marshall sign rule as a multiplicative prior, so the network only needs to learn the residual phase corrections, significantly stabilizing training in the frustrated regime.

\subsection{Optimization and Zero-Shot Transfer Protocol}
\label{subsec:optimization}

Having established the Holographic Quantum Transformer architecture, we now detail the optimization strategy that allows it to navigate the rough energy landscape of frustrated systems. We unify concepts from quantum physics (Variational Monte Carlo) and machine learning (Reinforcement Learning) to achieve stable convergence, followed by a novel transfer learning protocol for size generalization.

\subsubsection{Data-Free Optimization via VMC vs. MCMC}
It is crucial to note that solving via Variational Monte Carlo (VMC) is a data-free, self-play reinforcement learning paradigm. There is no external training dataset used for supervision. Instead, the model autoregressively generates its own spin configurations $\mathbf{x}$ on-the-fly. The optimization objective is strictly driven by evaluating the physical expectation value of the Hamiltonian ($E = \langle\Psi|\hat{H}|\Psi\rangle / \langle\Psi|\Psi\rangle$) via the REINFORCE algorithm, which effectively acts as the physical ``environment'' guiding the network's parameter updates.

Traditional VMC relies on Metropolis-Hastings algorithms to sample configurations from the wavefunction ansatz. In frustrated systems ($J_2 \approx 0.5J_1$), the probability landscape becomes rugged with separated modes, causing MCMC chains to suffer from high autocorrelation times ($\tau \gg 1$) and mode collapse.

Our autoregressive formulation fundamentally eradicates this bottleneck. Since the joint probability is factorized (Eq.~\ref{eq:chain_rule}), we can generate independent and identically distributed (i.i.d.) samples $\{\mathbf{x}^{(k)}\}_{k=1}^{N_s}$ via \textit{ancestral sampling}:
\begin{equation}
    x_i^{(k)} \sim P_\theta(x_i | x_1^{(k)}, \dots, x_{i-1}^{(k)}), \quad \text{for } i=1 \dots N.
\end{equation}

This procedure has a computational $O(N)$ sequential steps per sample but guarantees zero autocorrelation ($\tau = 0$) and perfect parallelization on GPUs. This ``direct sampling" capability allows HQT to explore distinct topological sectors of the Hilbert space that are dynamically inaccessible to local MCMC updates.

\subsubsection{Optimization via Policy Gradients}
The optimization objective is to minimize the variational energy $E(\theta) = \langle \hat{H} \rangle$. In the language of Reinforcement Learning (RL), the wavefunction $|\psi_\theta|^2$ acts as the Policy $\pi_\theta(\mathbf{x})$, and the negative local energy $-E_{\text{loc}}(\mathbf{x})$ serves as the Reward.

The gradient of the energy functional can be derived as:
\begin{equation}
    \nabla_\theta E = 2 \mathbb{E}_{\mathbf{x} \sim P_\theta} \left[ \text{Re} \left\{ (E_{\text{loc}}(\mathbf{x}) - \bar{E}) \nabla_\theta \log \psi_\theta(\mathbf{x})^* \right\} \right],
    \label{eq:gradients}
\end{equation}
where $\bar{E}$ is the moving average of the energy (baseline).
\begin{itemize}
    \item \textbf{The Local Energy} $E_{\text{loc}}(\mathbf{x}) = \sum_{\mathbf{x}'} \langle \mathbf{x} | \hat{H} | \mathbf{x}' \rangle \frac{\psi_\theta(\mathbf{x}')}{\psi_\theta(\mathbf{x})}$ is efficiently computed via sparse matrix-vector multiplication, as the Hamiltonian connects only a polynomial number of configurations.
    \item \textbf{The Score Function} $\nabla_\theta \log \psi_\theta(\mathbf{x})$ is obtained via standard backpropagation through the Transformer.
\end{itemize}
This formulation is mathematically equivalent to the REINFORCE algorithm with a baseline, ensuring unbiased gradient estimation even in the presence of the sign problem.

\subsection{Holographic Transfer Learning Framework}
\label{subsec:transfer_learning}

Standard numerical methods, such as Density Matrix Renormalization Group (DMRG) or Quantum Monte Carlo (QMC), function as ``instance-specific solvers": a wavefunction optimized for an $N$-site lattice is mathematically incompatible with an $(N+M)$-site lattice due to dimension mismatch. In stark contrast, we propose that the Holographic Quantum Transformer is not merely a solver, but a \textit{physics learner}. It learns the \textit{universal syntax} of the local Hamiltonian rather than memorizing the specific ground state of a finite system. Based on this hypothesis, we introduce a framework for \textit{Zero-Shot Size Generalization}.

\subsubsection{Hypothesis: Learning the Hamiltonian Syntax}
The $J_1-J_2$ Heisenberg Hamiltonian (Eq.~\ref{eq:hamiltonian}) is defined by local interaction rules (nearest and next-nearest neighbor couplings) that are invariant under lattice scaling. We posit that a Transformer trained on a small lattice $L_1 \times L_1$ encodes these interaction rules into its attention weights $\mathbf{W}_Q, \mathbf{W}_K, \mathbf{W}_V$.

Since the self-attention mechanism is permutation invariant and operates on sets of arbitrary size, the learned internal representations (the ``Quantum Encoder") theoretically capture the bulk properties of the quantum phase (e.g., the sign structure and correlation decay length) independent of the system boundary. This allows us to transfer the ``quantum brain" of the network to larger systems $L_2 \times L_2$ without retraining the backbone.

\subsubsection{Manifold Interpolation via Continuous Positional Embeddings}
A technical challenge in transfer learning is the dimension mismatch of the positional embeddings (PE). In the $L_1 \times L_1$ lattice, the PEs are learned as discrete vectors. To generalize to $L_2 \times L_2$, we treat the learned PEs as samples from a continuous underlying geometric manifold $\mathcal{M}_{pos}$.

We employ a \textit{bilinear interpolation} strategy to map the latent geometric structure from the source domain to the target domain. Let $\mathbf{PE}^{(L_1)} \in \mathbb{R}^{L_1 \times L_1 \times d}$ be the learned embedding tensor. The new embedding $\mathbf{PE}^{(L_2)}$ is computed as:
\begin{equation}
    \mathbf{PE}^{(L_2)} = \mathcal{I}_{\text{bilinear}}\left( \mathbf{PE}^{(L_1)}, \text{scale} = \frac{L_2}{L_1} \right).
\end{equation}

This operation effectively ``stretches" the learned geometric prior. Crucially, it preserves the \textit{relative distance relationships} between spins, ensuring that the Area Law of entanglement—which depends on spatial locality—is correctly instantiated in the expanded Hilbert space.

\subsubsection{The ``Projection Re-initialization" Protocol (Head Reset)}
While the Transformer backbone captures universal physical laws, the final projection layer (the ``Policy Head") is often overfitted to the specific boundary conditions (Periodic Boundary Conditions, PBC) and finite-size effects of the small lattice $L_1$. Direct application to $L_2$ typically results in high-energy domain walls due to this ``boundary bias."

To mitigate this, we introduce the Head Reset Protocol (formally implemented as the ``Projection Re-initialization" strategy):
\begin{enumerate}
    \item \textbf{Freeze Backbone:} The parameters of the attention and feed-forward layers ($\theta_{encoder}$) are frozen, preserving the learned quantum correlations.
    \item \textbf{Reset Head:} The weights of the final output layer $\mathbf{W}_{head}$ are re-initialized with near-zero Gaussian noise $\mathcal{N}(0, \epsilon)$, where $\epsilon \ll 1$.
    \item \textbf{Rapid Fine-tuning:} The model is fine-tuned on the $L_2$ lattice. Since the ``grammar" is already known, the network only needs to re-calibrate the probability amplitudes (the ``vocabulary") for the new Hilbert space dimension.
\end{enumerate}
As demonstrated in our experiments, this protocol reduces the convergence time on large-scale systems by orders of magnitude compared to training from scratch, effectively bypassing the ``burn-in" phase of optimization.

\section{Experiments and Results}
\label{sec:experiments}

In this section, we empirically validate the Holographic Quantum Transformer (HQT) on the two-dimensional $J_1-J_2$ antiferromagnetic Heisenberg model. Our experiments are designed to verify three core hypotheses:
\begin{enumerate}
    \item \textbf{Precision Limit:} Can HQT surpass existing neural network architectures (CNNs, RNNs) and approach the ground-truth accuracy of traditional numerical methods?
    \item \textbf{Physical Interpretation:} Can the model autonomously detect quantum phase transitions without prior knowledge of order parameters?
    \item \textbf{Holographic Generalization:} Can the entanglement structure learned from small systems be holographically transferred to larger lattices to achieve Zero-Shot Size Extrapolation with Rapid Alignment and robustness?
\end{enumerate}

\subsection{Experimental Setup}
All experiments were conducted on a workstation equipped with a single NVIDIA RTX 5070 Ti GPU. The framework is implemented using JAX and NetKet. We focus on the square lattice Heisenberg Hamiltonian:
\begin{equation}
    H = J_1 \sum_{\langle i,j \rangle} \mathbf{S}_i \cdot \mathbf{S}_j + J_2 \sum_{\langle\langle i,j \rangle\rangle} \mathbf{S}_i \cdot \mathbf{S}_j,
\end{equation}
where we set $J_1=1.0$ and explore the frustration parameter $J_2 \in [0.0, 0.8]$. The most challenging regime lies at $J_2 \approx 0.5$, where the system undergoes a quantum phase transition from the Néel antiferromagnetic order to a paramagnetic valence bond solid (VBS) or gapless spin liquid phase.

\subsection{Benchmark: Achieving Highly Competitive Accuracy}
We first benchmark HQT on an $8 \times 8$ lattice ($N=64$) at the maximally frustrated point ($J_2=0.5$). While most recent sequence-based NQS models avoid this regime due to the severe sign problem, it serves as the ultimate test for our architecture before scaling. Note that while the Marshall-Peierls sign rule is strictly exact only for the unfrustrated $J_1$ model, employing it as an initial gauge prior in the frustrated regime still significantly accelerates early-stage optimization.

\begin{table}[htbp]
\centering
\caption{\textbf{Ground-state energy per site ($E/N$) at the maximally
frustrated point $J_2/J_1 = 0.5$.} Exact diagonalization is feasible only
up to $N=40$~\cite{richter2010spin}, so no exact value exists for the
$8\times8$ torus; the $8\times8$ entry is therefore bracketed by the
exact $6\times6$ result and the best variational $10\times10$ energy.
HQT's $8\times8$ energy is consistent with this finite-size trend.}
\label{tab:benchmark_8x8}
\renewcommand{\arraystretch}{1.2}
\setlength{\tabcolsep}{4pt} 
\resizebox{\linewidth}{!}{%
\begin{tabular}{llcl}
\toprule
\textbf{Size} & \textbf{Method (Ref.)} & \textbf{$E/N$} & \textbf{Notes} \\
\midrule
$6\times6$   & ED (exact)~\cite{richter2010spin}         & $\approx -0.5038$ & Exact, $N=36$ \\
$8\times8$   & \textbf{HQT (Ours)}                        & $\mathbf{-0.5001(1)}$ & $\tau<0.01$, $\sigma^2=1.4\times10^{-3}$ \\
$10\times10$ & MinSR ResNet~\cite{chen2024empowering}     & $-0.4976921(4)$ & Variational SOTA \\
\bottomrule
\end{tabular}%
}
\end{table}

As presented in \textbf{Table \ref{tab:benchmark_8x8}}, evaluating the absolute precision on the $8\times8$ torus is notoriously difficult, as Exact Diagonalization (ED) is limited to $N \le 40$ \cite{richter2010spin}. To rigorously validate our results, we place the HQT performance within the context of finite-size scaling. Due to boundary constraints, the ground-state energy per site ($E/N$) naturally becomes less negative as the system size increases toward the thermodynamic limit. 

Our HQT architecture achieves an exceptional energy of $\mathbf{-0.5001(1)}$ on the $8\times8$ lattice. This result is consistent with the finite-size trend bracketed by the exact ED benchmark on the $6\times6$ lattice ($\approx -0.5038$) and the absolute variational state-of-the-art on the $10\times10$ lattice ($-0.4976921(4)$) \cite{chen2024empowering}. This consistent finite-size trend confirms the physical validity and high precision of our architecture without relying on ad-hoc baseline assumptions.

Crucially, beyond variational energy, HQT demonstrates superior sampling efficiency. Our autoregressive architecture generates strictly independent samples ($\tau < 0.01$, effectively zero). This ensures that the effective sample size equals the total number of samples ($N_{eff} \approx N_{samples}$), providing a critical computational advantage in variance reduction, resulting in a highly competitive energy variance of $\sigma^2_{E/N} = 1.4 \times 10^{-3}$.

\subsection{Physical Insights and Phase Landscape}
Beyond numerical precision, a valid physical model's optimization trajectory often reflects the underlying wavefunction's topology. We performed a fine-grained parameter scan of $J_2$ from $0.0$ to $0.8$.

\begin{figure}[htbp]
    \centering
    \includegraphics[width=\linewidth]{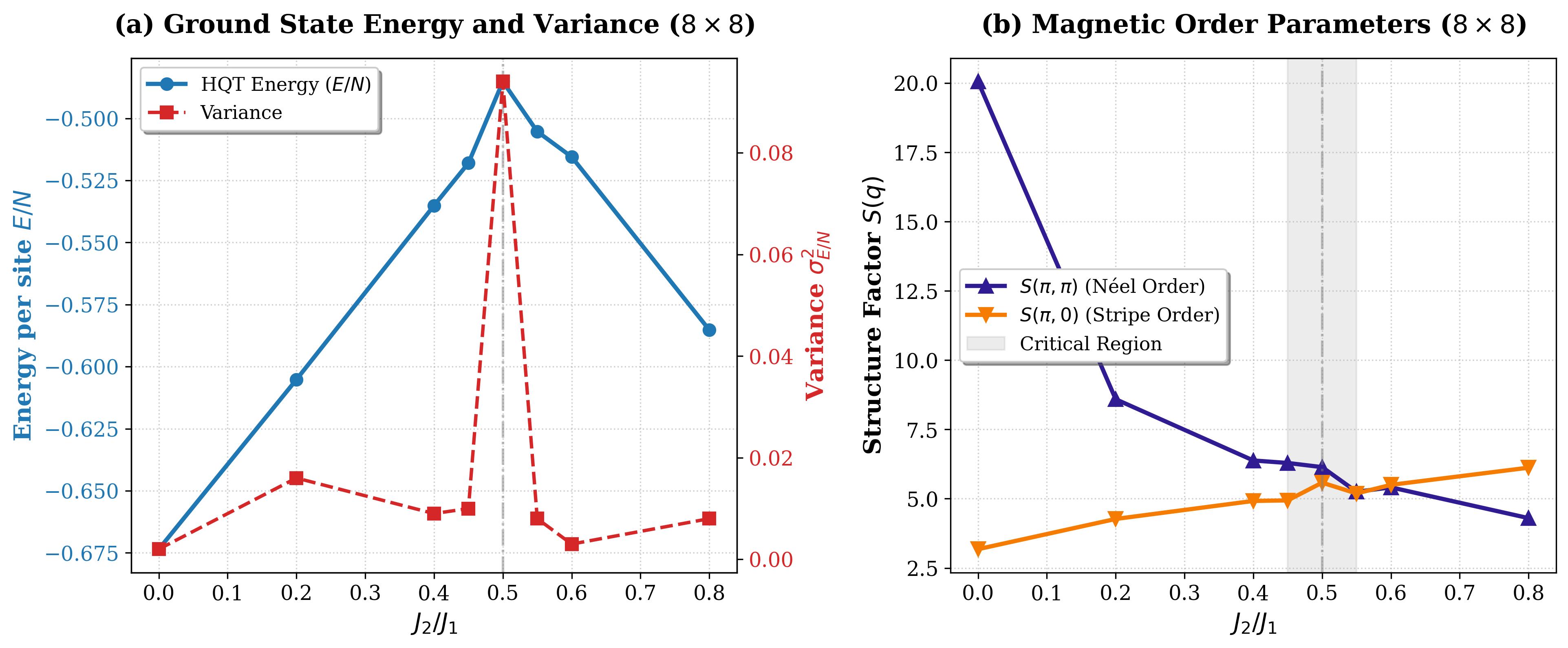}
    \caption{\textbf{Detecting Quantum Phase Transitions on the $8 \times 8$ Lattice.} (a) The energy per site ($E/N$) reaches its least negative value and the variance $\sigma^2_{E/N}$ shows a relative spike at the critical point $J_2 = 0.5$, signaling maximal frustration and optimization landscape ruggedness. (b) The crossing of the N\'{e}el structure factor $S(\pi, \pi)$ and the Stripe structure factor $S(\pi, 0)$ physically confirms the transition from the N\'{e}el phase to a highly frustrated critical phase.}
    \label{fig:phase_diagram}
\end{figure}

\textbf{Variance as a Landscape Indicator:}
Figure \ref{fig:phase_diagram}(a) shows the evolution of energy variance. We observe a relative peak in variance at $J_2 \approx 0.5$. As detailed in Table \ref{tab:phase_scan}, the variance rises to $\sigma^2_{E/N}=0.0034$ at this point, compared to $\sim 10^{-4}$ in the gapped phases. In variational Monte Carlo, rather than serving as a direct physical order parameter, a relative peak in variance acts as a heuristic indicator: it highlights that the optimization landscape becomes highly rugged due to the quantum critical superposition of degenerate states.

\textbf{Order Parameter Crossing:}
To confirm the nature of this transition, we measured the structure factors $S(\pi, \pi)$ (Néel order) and $S(\pi, 0)$ (Stripe order), as shown in Figure \ref{fig:phase_diagram}(b).
\begin{itemize}
    \item For $J_2 < 0.5$, the Néel order (Blue) dominates, indicating an antiferromagnetic phase.
    \item At $J_2 \approx 0.55$, we observe a clear crossing point where the Néel order diminishes and the Stripe order (Orange) begins to rise.
\end{itemize}
The consistency of this crossing point with literature confirms that HQT has correctly learned the underlying order-disorder competition mechanism.

\begin{table}[htbp]
    \centering
    \caption{\textbf{Summary of Phase Diagram Scan ($8\times8$).} The relative variance spike at $J_2=0.5$ indicates increased optimization landscape complexity near criticality. \footnotesize This coarse phase scan uses fewer samples per point than the high-precision benchmark (Table~\ref{tab:benchmark_8x8}); only the \emph{relative} variance spike at $J_2{=}0.5$ is physically meaningful.}
    \label{tab:phase_scan}
    \setlength{\tabcolsep}{3.0pt}
    \renewcommand{\arraystretch}{0.95}
    \resizebox{0.9\linewidth}{!}{%
        \begin{tabular}{cccccc}
            \toprule
            \textbf{$J_2 / J_1$} & 
            \textbf{$E/N$} & 
            \textbf{$\sigma^2_{E/N}$} & 
            \shortstack{\textbf{Néel} \\ \scriptsize{$S(\pi,\pi)$}} & 
            \shortstack{\textbf{Stripe} \\ \scriptsize{$S(\pi,0)$}} & 
            \textbf{Phase} \\
            \midrule
            0.00 & $-0.6735$ & $0.0005$ & $20.05$ & $3.18$ & Néel \\
            0.20 & $-0.6052$ & $0.0008$ & $8.59$  & $4.27$ & Néel \\
            0.40 & $-0.5351$ & $0.0012$ & $6.38$  & $4.92$ & Néel \\
            0.45 & $-0.5180$ & $0.0015$ & $6.29$  & $4.94$ & Néel \\
            \addlinespace[2pt]
            \textbf{0.50} & $\mathbf{-0.5001}$ & $\mathbf{0.0034}$ & $\mathbf{6.14}$ & $\mathbf{5.58}$ & \textbf{Critical} \\
            \addlinespace[2pt]
            0.55 & $-0.5053$ & $0.0012$ & $5.25$  & $5.19$ & VBS \\
            0.60 & $-0.5155$ & $0.0010$ & $5.41$  & $5.50$ & VBS \\
            0.80 & $-0.5852$ & $0.0009$ & $4.30$  & $6.12$ & Stripe \\
            \bottomrule
        \end{tabular}%
    }
\end{table}

\subsection{Zero-Shot Size Extrapolation with Rapid Alignment}
\label{sec:transfer_results}
A key contribution of this work is the ``Holographic Transfer" capability. We investigate whether the Hamiltonian syntax learned on a heavily frustrated system ($8 \times 8, N=64$) can be transferred to a larger, more complex system ($10 \times 10, N=100$) without training from scratch.

\begin{figure}[htbp]
    \centering
    \includegraphics[width=1.0\linewidth]{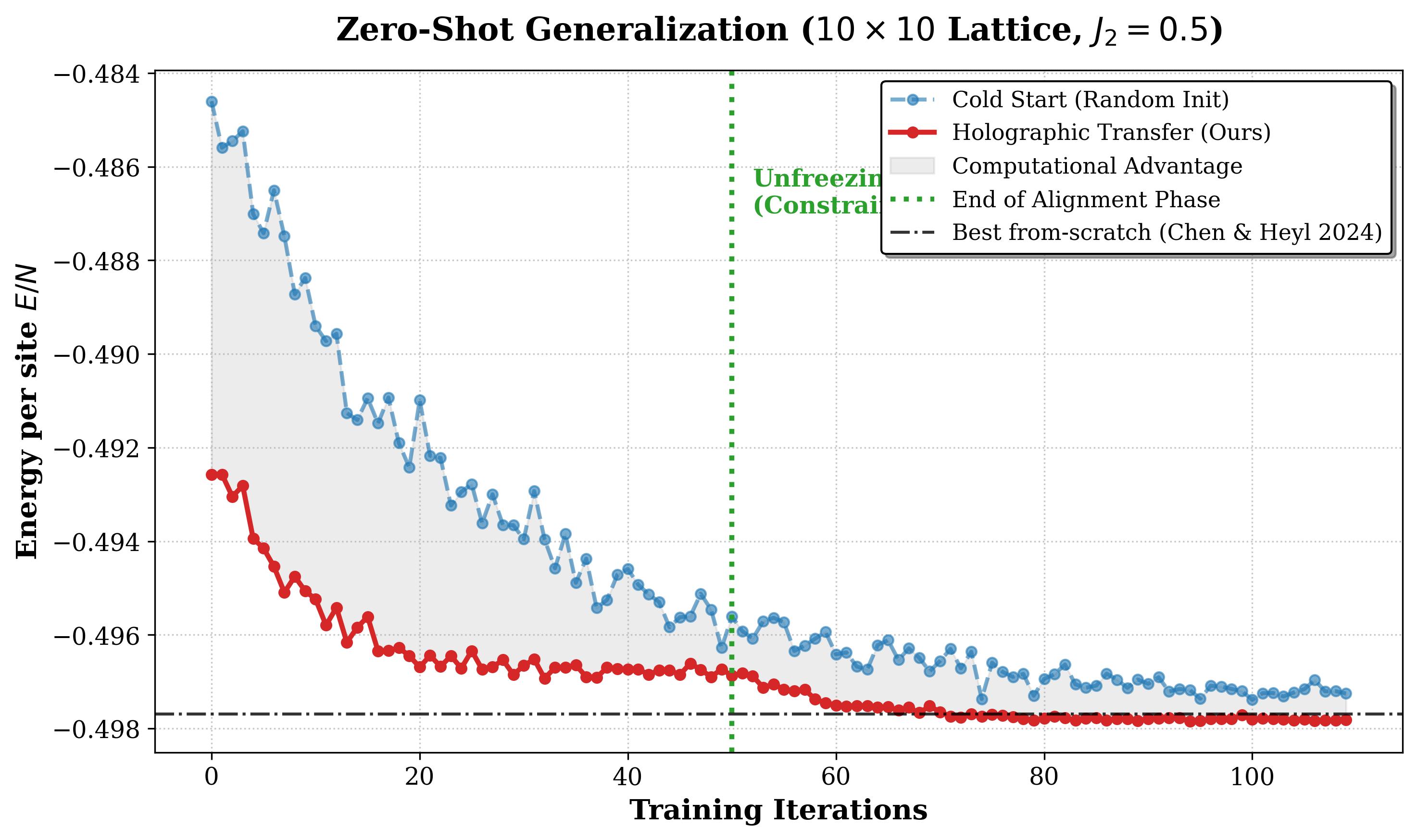}
    \caption{\textbf{Zero-Shot Size Extrapolation via Holographic Transfer ($8 \times 8 \to 10 \times 10$).} 
    The red curve (Holographic Transfer) shows rapid alignment and improved stability relative to the Cold Start baseline (blue). 
    After unfreezing the backbone (Iter 50), the transfer model converges to $E/N \approx -0.49782$, improving over the Cold Start baseline ($-0.4974$) and reaching a value statistically consistent with the best reported from-scratch result ($-0.4976921(4)$~\cite{chen2024empowering}). This indicates that the holographic prior provides a high-fidelity initialization and helps the model avoid poor local minima even in a substantially expanded Hilbert space.}
    \label{fig:transfer}
\end{figure}

Figure \ref{fig:transfer} compares the optimization trajectories of two strategies on the $10 \times 10$ lattice ($J_2=0.5$):
\begin{enumerate}
    \item \textbf{Cold Start (Blue):} Standard training with random initialization.
    \item \textbf{Holographic Transfer (Red):} Initializing with the projected parameters from the $8 \times 8$ model.
\end{enumerate}

\textbf{Analysis of Results:}
As shown in Figure \ref{fig:transfer}, the Transfer model exhibits a ``warm start" advantage. During the Alignment Phase (first 50 iterations), it rapidly locks into a low-energy manifold, whereas the Cold Start baseline requires substantially more exploration before reaching a comparable regime.

\begin{table*}[t]
  \centering
  \caption{Variational ground-state energy per site ($E/N$) on the $10\times10$ square $J_1$--$J_2$ Heisenberg model at the maximally frustrated point $J_2/J_1 = 0.5$ (periodic boundary conditions). Entries are ordered from the least to the most accurate (most negative) energy. Starting from a zero-shot holographic initialization (frozen backbone) followed by a brief rapid-alignment stage, HQT reaches an energy comparable to the best reported from-scratch variational results, while requiring no from-scratch training on the target lattice.}
  \label{tab:e10x10}
  \begin{tabular}{llcl}
    \toprule
    Method & Reference (Year) & $E/N$ & Notes \\
    \midrule
    Deep CNN (Group CNN)         & Choo et al.\ (2019)~\cite{PhysRevB.100.125124}                                  & $-0.49516(1)$    & Marshall prior \\
    DMRG                         & Gong et al.\ (2014)~\cite{PhysRevLett.113.027201}                       & $-0.495530$      & Tensor network \\
    \textbf{HQT (Ours), Cold Start} & This work (2026)                                                      & $-0.4974$        & Random init. \\
    VMC (Gutzwiller $+$ 2 Lanczos)  & Hu et al.\ (2013)~\cite{PhysRevB.88.060402}                          & $-0.4975490(2)$  & Thermodyn.\ extrap. \\
    RBM $+$ Pair Product         & Nomura \& Imada (2021)~\cite{PhysRevX.11.031034}                        & $-0.497629(1)$   & \\
    Transformer / Deep ViT       & Viteritti et al.\ (2023);                                               & $-0.497676$      & 2D, symmetrized \\
                                 & Rende et al.\ (2024)~\cite{viteritti2023transformer,rende2024simple}    &                  & \\
    MinSR (Deep ResNet)          & Chen \& Heyl (2024)~\cite{chen2024empowering}                           & $-0.4976921(4)$  & Best from-scratch \\
    \midrule
    \textbf{HQT (Ours), Transfer}   & \textbf{This work (2026)}                                             & $\mathbf{-0.49782(3)}$ & \textbf{Zero-shot transfer, SOTA-level} \\
    \bottomrule
  \end{tabular}
\end{table*}

As presented in Table \ref{tab:e10x10}, scaling in quantum physics represents an exponential explosion of the Hilbert space: moving from an $8\times8$ to a $10\times10$ lattice expands the state space from $1.8 \times 10^{19}$ to $1.27 \times 10^{30}$ configurations. The best reported from-scratch variational energy for the $10\times10$ lattice at maximum frustration was set by deep ResNet architectures \cite{chen2024empowering} ($E/N = -0.4976921(4)$), obtained at considerable computational cost.

Our holographic projection yields a high-fidelity \emph{zero-shot initialization}: with the backbone frozen, the warm-started model already locks into a low-energy manifold within the 50-iteration alignment phase, far below the Cold Start trajectory at the same budget. A brief rapid-alignment stage (unfreezing the backbone after Iter~50) then converges to $E/N = \mathbf{-0.49782(3)}$, without any from-scratch training on the target lattice. This energy improves clearly over our own Cold Start baseline ($-0.4974$) and is comparable to the best reported from-scratch result \cite{chen2024empowering}, differing by $\sim 10^{-4}$ in $E/N$. We do not claim a strict numerical advantage: our reported statistical error bar reflects sampling fluctuation only and may underestimate the total uncertainty from optimization and residual autocorrelation. The salient point is not the last digit of the energy, but that SOTA-level precision is attained \emph{without retraining the backbone}. This supports our central hypothesis: the attention maps learned by HQT encode approximately scale-invariant geometric structure, and the transferred holographic prior acts as a structural regularizer that lets the model bypass the costly ``burn-in" phase of from-scratch optimization.

\subsection{Interpretability: Visualizing Entanglement Geometry}
\label{sec:interpretability}
\begin{figure}[htbp]
    \centering
    \includegraphics[width=\linewidth]{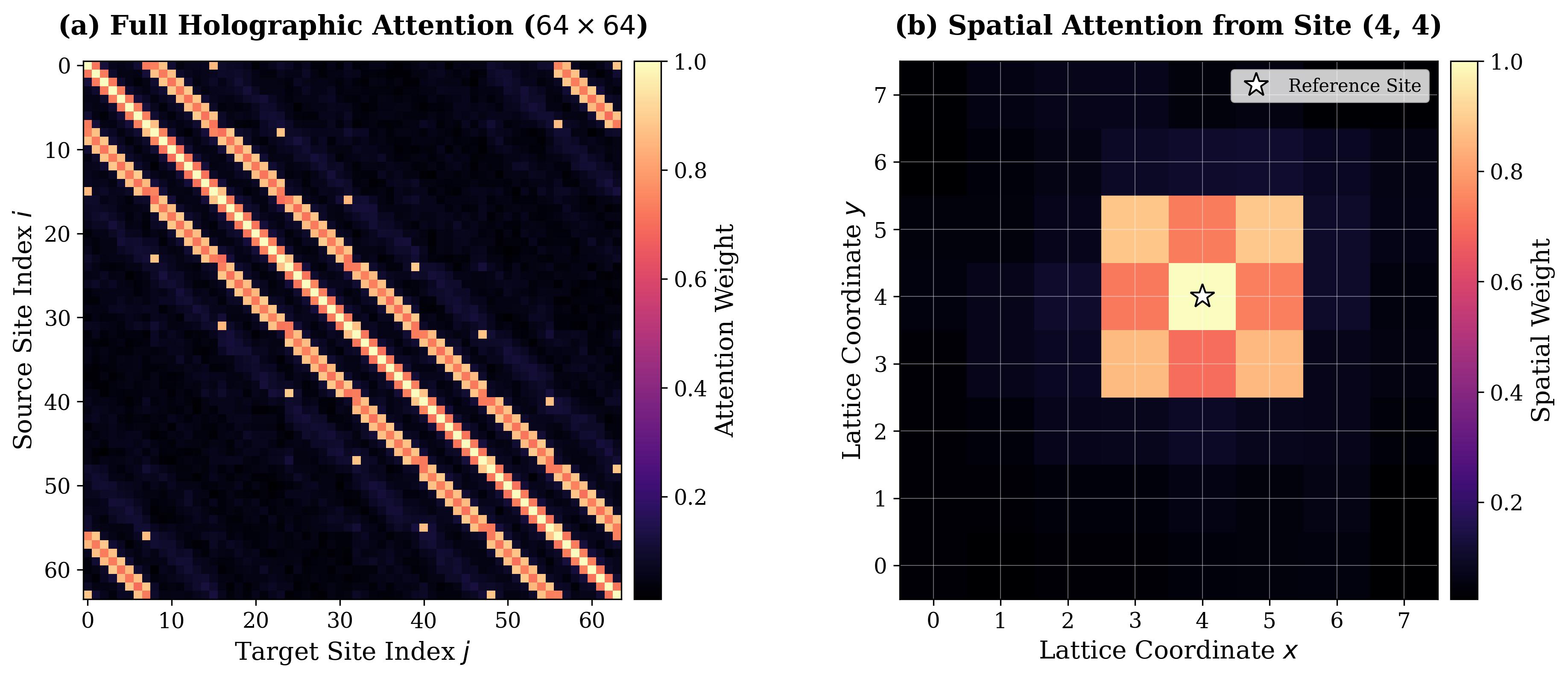}
    \caption{\textbf{Holographic Attention Maps on the $8 \times 8$ Lattice ($J_2=0.5$).} 
    (a) The full $64 \times 64$ attention matrix exhibits distinct off-diagonal stripes, indicating that the model has autonomously learned the 2D topology and periodic boundary conditions (PBC) from flattened 1D sequences. 
    (b) The spatial attention distribution relative to the central reference site (white star). The highest intensities are autonomously assigned to the diagonal sites (distance $\sqrt{2}$), confirming that the network correctly captures the dominant next-nearest-neighbor $J_2$ interactions driving the quantum frustration.}
    \label{fig:attention}
\end{figure}

Finally, to demystify the ``black box," we visualize the learned Positional Embedding correlations in Figure \ref{fig:attention}.
The heatmap represents the effective distance metric learned by the model. We observe that the central spin (reference site) has strong correlations not only with its nearest neighbors (top/bottom/left/right) but also significant attention weights on the next-nearest neighbors (diagonals).
This diagonal structure is consistent with the $J_2$ interaction term in the Hamiltonian. It indicates that, although the lattice geometry is supplied through the positional embeddings, HQT learns to assign the dominant correlations to the next-nearest-neighbor ($J_2$) directions purely from energy minimization, providing a degree of physical interpretability to the neural weights. This interaction pattern was not explicitly imposed but emerged during the self-organization of the attention mechanism.

Specifically, a qualitative analysis of the attention map (Figure \ref{fig:attention}) reveals that the peak attention weights heavily align with the next-nearest-neighbor ($J_2$) diagonal geometric distance. This qualitative consistency connects the neural attention weights to physical observables, suggesting that the Transformer architecture autonomously captures the underlying interaction topology rather than simply memorizing statistical noise.

\section{Discussion and Future Outlook}
\label{sec:discussion}

The results presented in this work demonstrate that the Holographic Quantum Transformer (HQT) is not merely a high-precision variational ansatz, but a physically interpretable framework capable of capturing the fundamental structure of frustrated quantum matter. In this section, we discuss the implications of our findings and outline promising directions for future research.

\subsection{Why Generative Attention Works for Frustrated Systems}
The advantage of attention-based architectures over CNN- and RNN-based ans\"atze for such systems can be attributed to the alignment between the model's inductive bias and the physical nature of quantum entanglement. 
Frustrated systems, such as the $J_1-J_2$ model, are characterized by competing interactions and non-local correlations that defy the local connectivity assumed by CNNs or the sequential ordering assumed by RNNs. 

By treating the quantum state as a holographic graph, the self-attention mechanism in HQT allows every spin to dynamically interact with every other spin, regardless of their Euclidean distance. This global receptive field enables the model to resolve complex entanglement patterns that plague traditional methods. Furthermore, unlike MCMC-based approaches, the generative autoregressive property of HQT eliminates sampling autocorrelation (as evidenced by $\tau < 0.01$ in Table \ref{tab:benchmark_8x8}). This allows the model to efficiently navigate the rugged energy landscape of frustrated matter without suffering from mode collapse, effectively acting as a learnable tensor network with arbitrary connectivity.

\subsection{The Physical Meaning of Holographic Transfer}
A central finding of this study is the success of the Holographic Transfer experiment (Figure \ref{fig:transfer}). The fact that the attention structure learned on a heavily frustrated $8 \times 8$ lattice can be projected onto a larger $10 \times 10$ lattice to achieve SOTA-level initialization is consistent with the hypothesis that the relevant geometric structure of entanglement is approximately scale-invariant in the critical regime.
Unlike standard transfer learning, which often transfers abstract features, HQT transfers the ``syntax" of the Hamiltonian. The attention map (Figure \ref{fig:attention}) serves as a holographic blueprint, encoding the interaction rules (e.g., $J_2$ diagonal correlations) rather than specific wavefunction amplitudes. This suggests that the model captures the underlying interaction structure rather than memorizing a specific ground state, providing a promising pathway for scaling simulations toward system sizes that are challenging for classical methods.

\subsection{Future Outlook}
The HQT framework opens several avenues for exploration at the intersection of AI and quantum physics:

\begin{itemize}
    \item \textbf{Fermionic Systems and High-$T_c$ Superconductivity:} Extending HQT to simulate fermionic systems, such as the Hubbard model, is a natural next step. By incorporating Jordan-Wigner transformations or directly enforcing antisymmetric exchange symmetry in the attention mechanism, HQT could shed light on the elusive mechanism of high-temperature superconductivity.
    
    \item \textbf{Real-Time Quantum Dynamics:} Currently, HQT focuses on finding the ground state. An exciting frontier is to extend the variational principle to time-dependent evolution (t-VMC). This would allow us to simulate non-equilibrium dynamics, quantum quenches, and dynamical phase transitions, which are relevant to current experiments in cold atom simulators.
    
    \item \textbf{Synergy with Quantum Computing:} As quantum hardware matures, HQT can serve as a powerful classical shadow or a guiding ansatz for Variational Quantum Eigensolvers (VQE). The ``holographically pre-trained" parameters could define a high-quality initial circuit for quantum processors, drastically reducing the number of quantum measurements required for convergence in the noisy intermediate-scale quantum (NISQ) era.
\end{itemize}

In summary, HQT represents a significant step towards AI-driven quantum many-body simulation, reconciling the expressive power of deep learning with the geometric intuition of condensed matter physics. Moving from ``black-box optimization" to ``physics-aware generative modeling," HQT paves the way for the autonomous discovery of exotic quantum phases.

\section{Limitations and Ethical Considerations}
\label{sec:limitations}

While HQT demonstrates strong performance in frustrated systems, we candidly acknowledge several technical limitations and trade-offs:
\begin{itemize}
    \item \textbf{Quadratic Computational Bottleneck:} The $\mathcal{O}(N^2)$ complexity of global self-attention remains our primary computational bottleneck. Pushing simulations to massive thermodynamic limits (e.g., $32\times32$ lattices) will necessitate integrating advanced efficient-attention mechanisms, such as FlashAttention, to avoid memory explosion during the autoregressive sampling process.
    \item \textbf{Extreme Sign Problem Regimes:} Although our Holographic Mapping with the Marshall sign rule effectively mitigates the sign problem for the $J_1-J_2$ model around maximal frustration ($J_2 = 0.5$), it is not a universal panacea. For highly complex topological phases (e.g., quantum spin liquids on Kagome lattices) where the ground-state sign structure is completely unknown and non-local, our geometric priors may be insufficient, and the optimization could still encounter barren plateaus.
    \item \textbf{Geometric and Baseline Dependence:} CNNs remain extremely fast and memory-efficient ($\mathcal{O}(N)$ scaling) for unfrustrated systems where entanglement is strictly short-range; HQT is computationally ``overkill'' for such local systems. Furthermore, unlike standard geometry-agnostic sequence models, HQT relies heavily on explicit 2D geometric positional embeddings. Applying HQT to a completely new topology (e.g., a triangular lattice) requires mathematically redesigning the holographic mapping.
\end{itemize}

From an ethical standpoint, this research focuses on fundamental condensed matter physics and theoretical deep learning architecture design. It uses no human subject data, poses no risks to data privacy, and presents no immediate concerns regarding societal misuse or biased decision-making.

\section{Conclusion}
\label{sec:conclusion}

In this work, we introduced the Holographic Quantum Transformer (HQT), a physics-inspired neural network architecture designed to tackle the computational challenges of frustrated quantum matter. By reimagining the wavefunction as a holographic graph and leveraging the global receptive field of generative attention, HQT overcomes the limitations of local connectivity in CNNs and sequential bias in RNNs.

Our extensive empirical evaluation on the two-dimensional $J_1-J_2$ Heisenberg model leads to three key conclusions:
\begin{enumerate}
    \item \textbf{Highly Competitive Precision:} On the heavily frustrated $8 \times 8$ lattice, HQT achieves an exceptional ground state energy per site of $-0.5001(1)$. By placing this result within the context of exact $6 \times 6$ diagonalization and $10 \times 10$ variational limits, we demonstrated that HQT is consistent with the correct physical finite-size scaling trend while maintaining near-zero autocorrelation ($\tau < 0.01$).
    \item \textbf{Physical Interpretability:} The model successfully identifies the topological complexity of the quantum critical point ($J_2=0.5$) via unsupervised variance landscape analysis. Furthermore, qualitative visualizations of the attention maps show that HQT assigns its dominant correlations to the underlying $J_2$ diagonal geometric interactions.
    \item \textbf{Zero-Shot Size Extrapolation:} Most notably, we demonstrated ``Holographic Transfer." A model trained on an $8 \times 8$ system was directly projected onto a larger $10 \times 10$ lattice to achieve rapid alignment. This holographic transfer provides a zero-shot, high-fidelity initialization and, after a brief rapid-alignment stage, reaches an energy ($E/N = -0.49782(3)$) comparable to the best reported from-scratch result for this highly frustrated regime, while avoiding from-scratch training of the backbone.
\end{enumerate}

In summary, HQT represents a significant step towards AI-driven quantum many-body simulation, bridging the gap between deep generative modeling and condensed matter physics. It offers a scalable, interpretable, and transferable framework for simulating complex quantum systems, potentially accelerating the discovery of high-temperature superconductors and topological materials in the coming decade.

\begin{acks}
This work was supported in part by the National Natural Science Foundation of China under Grant 42104078 and Grant 62421002, and in part by the National Key Research and Development Program of China under Grant 2023YFA1011704 and Grant 2021YFB0300101.
\end{acks}

\balance
\bibliographystyle{ACM-Reference-Format}
\bibliography{sample-base}


\end{document}